# Artificial intelligence for objective assessment of acrobatic movements: How to apply machine learning for identifying tumbling elements in cheer sports

Sophia Wesely[1], Ella Hofer [2], Robin Curth [1,3], Shyam Paryani [4], Nicole Mills [5], Olaf Ueberschär [2,6,#,*], Julia Westermayr [1,3,#,*]

[1] Leipzig University, Faculty of Chemistry and Mineralogy, Institute of Physical and Theoretical Chemistry, Leipzig, Germany
[2] Magdeburg-Stendal University of Applied Sciences, Department of Engineering and Industrial Design, Magdeburg, Germany
[3] Center for Scalable Data Analytics and Artificial Intelligence, Dresden/Leipzig, Germany
[4] University of North Florida, Brooks College of Health, Jacksonville, FL, USA
[5] University of North Florida, Athletics Department, Jacksonville, FL, USA
[6] Institute for Applied Training Science, Leipzig, Germany
\# These two senior authors contributed equally.
\* Correspondence: OUe: olaf.ueberschaer@h2.de, JW: julia.westermayr@uni-leipzig.de

**Abstract**

Over the past four decades, cheerleading has evolved from a sideline activity at major sporting events into a professional, competitive sport with growing global popularity. Evaluating tumbling elements in cheerleading relies on both objective measures and subjective judgments, such as difficulty and execution quality. However, the complexity of tumbling—encompassing team synchronicity, ground interactions, choreography, and artistic expression—makes objective assessment challenging. Artificial intelligence (AI) has revolutionized various scientific fields and industries through precise data-driven analyses, yet their application in acrobatic sports remains limited despite significant potential for enhancing performance evaluation and coaching. This study investigates the feasibility of using an AI-based approach with data from a single inertial measurement unit to accurately identify and objectively assess tumbling elements in standard cheerleading routines. A sample of 16 participants (13 females, 3 males) from a Division I collegiate cheerleading team wore a single inertial measurement unit at the dorsal pelvis. Over a 4-week seasonal preparation period, 1102 tumbling elements were recorded during regular practice sessions. Using triaxial accelerations and rotational speeds, various ML algorithms were employed to classify and evaluate the execution of tumbling manoeuvres. Results indicate that certain machine learning models can effectively identify different tumbling elements despite inter-individual variability and data noise, achieving high accuracy. These findings demonstrate the significant potential for integrating AI-driven assessments into cheerleading and other acrobatic sports, providing objective metrics that complement traditional judging methods.

## 1. Introduction

Over the past few decades, cheerleading has evolved from the sidelines of major team sport events to a professional competitive sport of its own right, and now belongs to one of the most popular sports in the USA, with more than 3 million participants annually, and increasing worldwide popularity [1,2]. On the modern competitive level, the choreographed routines of cheerleading combine elements from gymnastics-like tumbling, stunts and pyramid acrobatics, jumps, dancing and cheering [2,3]. Judges' scoring is based on several objective and subjective categories comprising the difficulty and quality of execution of standing and running tumbling and individual, pair or group stunts, crowd interaction, expression, and motion technique [3]. Although tumbling involves many gymnastic-like manoeuvres, its character and execution in cheerleading differ substantially due to the aspect of team synchronicity, different ground properties and their alignment to overall choreography and artistic expression. Therefore, an objective assessment of the quality of execution of tumbling elements may prove challenging to the judges in competition and coaches during exercise given the fast, parallel execution of several athletes and their integration into a choreographic composition.

With the advent of artificial intelligence in several areas of research and aspects of daily life, e.g. voice assistants like Siri, autonomous driving technologies, or diagnostics in medicine [4-6], the potential for AI to also assist in sports biomechanics and exercise science has garnered considerable interest [7-12]. In the realm of sports science, AI and machine learning (ML), a subset of AI using statistical algorithms to learn from data [13], are being increasingly applied in biomechanical gait analysis, performance optimization, injury prevention and team tactics [9,14,15]. These technologies, often used in combination with wearables that monitor heart rate, sleep quality, or other daily activities, offer the capability to process and analyse large volumes of data with high precision, enabling more objective and consistent assessments compared to traditional methods [16]. Thus, AI and ML can be seen as an assistance to provide support for training by means of data analysis [17].

Despite these promising advancements in AI applications within sports, their utilization in acrobatic disciplines such as cheerleading remains limited. One study focuses on using ML in teaching cheerleading [18]. However, the inherent complexity of cheerleading tumbling, which involves dynamic movements, precise timing, and coordination among team members, presents both challenges and opportunities for AI-driven assessment. ML algorithms, particularly those that are adept at handling time-series data from inertial measurement units (IMUs), offer a promising approach to objectively quantify and evaluate tumbling elements. By capturing detailed motion data, these algorithms may be capable of identifying patterns and nuances in athletes' performances that may be difficult to discern through human observation alone [19].

However, the application of ML in cheerleading is not without its challenges. Due to the inherent dependence of each element's execution on the individual athlete's unique biomechanics and technique, there is a risk of the "Clever Hans" effect [20], where models learn to recognize specific individuals rather than the intended performance characteristics. This can lead to models that perform well on training data but fail to generalize to new athletes or different performance contexts. Overcoming the limitations associated with individual variability and ensuring model generalizability can significantly enhance the objectivity and precision of performance evaluations. Thus, addressing this issue requires careful model training, validation, and the incorporation of diverse datasets to ensure that ML algorithms can reliably assess tumbling elements across a broad range of performers [21].

In spite of these challenges, the potential benefits of integrating ML into the assessment of cheerleading are substantial. In this study, we seek to bridge the gap between ML and cheerleading by providing initial insights into the possibilities of using ML to support the assessment of cheerleading and tumbling elements. In particular, we investigate the feasibility of an ML-based approach that uses data from a single IMU to accurately identify tumbling elements in cheerleading routines. Using triaxial accelerations and rotational velocities, different ML algorithms are employed to classify and evaluate the execution of tumbling manoeuvres. Our analysis focuses on optimizing data processing techniques to improve ML prediction capabilities and ensure that the models can effectively handle the complexity of cheerleading performances. The results show that ML models can generalize and accurately predict tumbling elements, while generalizing to athletes never seen by the model. This integration into cheerleading has the potential to provide objective metrics that complement traditional scoring methods, thereby improving both competition evaluations and training methods as predictions fail in cases where manoeuvres are not executed cleanly. Our results show that ML has the potential to not only support fairer and more consistent scoring in the future, but also provides detailed feedback for athletes and coaches facilitating targeted training and performance improvements.

## 2. Materials and Methods

### 2.1. Subjects

A total of 16 well-trained collegiate cheerleaders (13 females, 3 males) from the coed cheerleading team of the University of North Florida, Jacksonville, FL, USA, took part in this study. Their age was 19.9±1.0 (18–22) years, they had a mean stature of 1.65±0.10 m (1.42–1.85 m) and a mean body mass of 65.3 ± 13.5 kg (43.1–94.3 kg). The cohort consisted of 6 female flyers and 15 mixed-sex bases (7 female, 3 male). All subjects reported to have been of injury for at least one month prior to the measurement. The team's highly experienced coach instructed each session and was present during the entire measurement time. The study was conducted in accordance with the Declaration of Helsinki, and approved by both the Ethics Committee of the Department of Engineering and Industrial Design of the Magdeburg-Stendal University of Applied Sciences (protocol code EKIWID-2025-02-001EH, date 2025-02-07) and the Institutional Review Board of the University of North Florida (protocol code IRB#1970666-2, date 2023-01-18). All participants gave written informed consent to their participation.

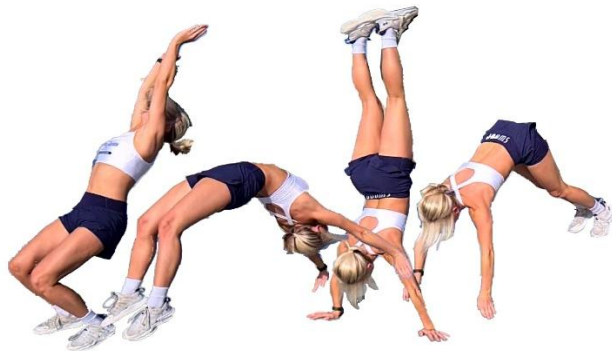

**Figure 1.** Chronophotography example of a back handspring executed by one of the study's cohort subjects. The IMU is worn under the pants approximately in the centre of the "Swoop" slogan print.

### 2.2. Tumbling Elements

During the team's three-month seasonal preparation period from September to November, a total of 1102 tumbling elements of six different types conducted by the 16 athletes were captured. The tumbling elements included were *Back Full (*BF*,* backflip with a full 360° twist), *Back Handspring* (BHS, the athlete jumps backwards into a reverse rotation about the transverse axis through a handstand position, blocks with the hands by shifting weight to the arms and pushes off with the shoulders to land back on the feet to complete the turn, Figure 1); *Back Layout* (BL, backflip performed with a fully extended to slightly hollow position in the air*), Back Tuck* (BT*,* backflip performed in with the knees being pulled towards the chest while the hip is flexed, creating a tuck position in the air), *Front Walkover* (FW, tumbling element in which the athlete moves through a handstand into a split-legged bridge position and then stands up smoothly), *Round Off* (RO, tumbling element in which the athlete supports the body weight with their arms while rotating sideways through an inverted position and landing on both feet on the ground at the same time, facing the direction from which they came) [3]. The distribution of these different tumbling elements is depicted in Figure 2. For each tumbling element, at least two different athletes performed the manoeuvre.

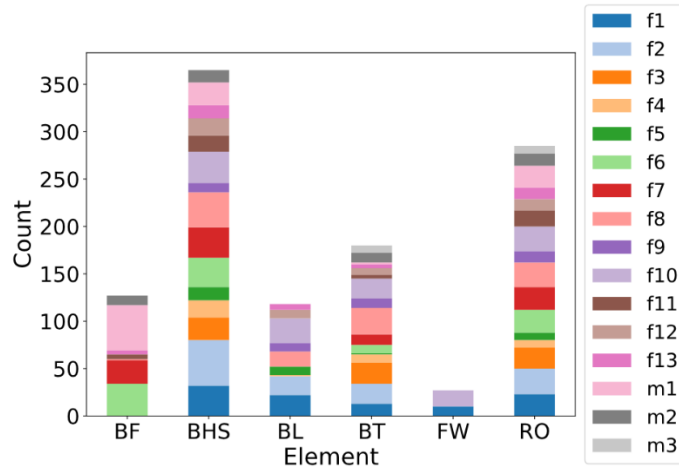

**Figure 2.** Distribution of tumbling elements performed by the 16 athletes (females f1–f13, males m1–m3) in the study. The six tumbling elements are Back Full (BF), Back Handspring (BHS), Back Layout (BL), Back Tuck (BT), Front Walkover (FW) and Round Off (RO).

*2.3. Data acquisition*

Single wireless IMUs (Xsens MTw Awinda, Movella Inc., Enschede, The Netherlands) combining tri-axial accelerometers (±16 *g*), gyroscopes (±2000°/s), and magnetometers (±800 µT) were placed in the athletes' lumbar regions near the first sacral vertebra (S1) using Velcro straps (Figure 3) [22]. For particularly slim athletes, the IMU position was further secured the tight-fitting shorts worn over it. The internal data sampling rate was 1000 Hz for the accelerometers and gyroscopes, while the output data rate after sensor fusion was typically 120 Hz, and in one session, due to a higher number of athletes, 100 Hz [23]. The data were transmitted from the IMU to a receiving base station via a proprietary wireless protocol of the manufacturer using the 2.4 GHz band with a maximum free space range of 50 m. The actual maximum distance between the IMU and the base station that occurred during the motion capture sessions was less than 10 m. The IMU was initialized according to the manufacturer's recommendations (i.e., proper temperature acclimatization period to adapt to changed indoor or outdoor conditions, sensor self-initialization in rest, sensor heading reset based on the ENU coordinate axis definition, etc.) before they were attached to the athlete [23]. The type of IMU utilised has been widely used and validated for a wide range of motion analysis purposes in human sports science [22-25].

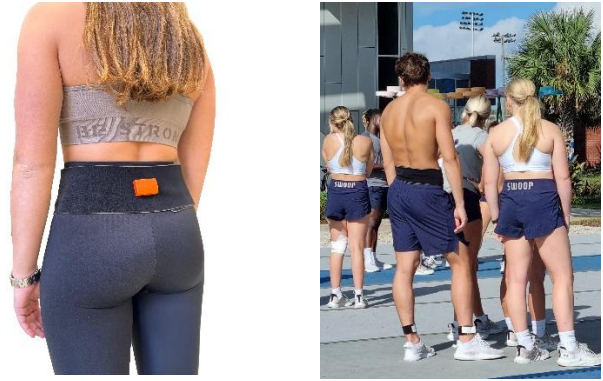

**Figure 3.** Utilised inertial measurement unit (IMU). **a)** The Xsens MTw Awinda IMU (orange) is placed in the athlete's lumbar region near the S1 vertebra using a comfortable, wide Velcro strap (black) put around the athlete's waist. b) The rest of Velcro strap is then put on top of the IMU to further fix its position, as can be seen for the shirtless male athlete in the front.

*2.4. Computational details:*

2.4.1. Data preprocessing and unsupervised learning analysis

To make the data machine readable, the raw data had to be pre-processed. Therefore, each element that was performed less than 10 times was removed from the data set in order to ensure a good split between training, validation, and test set for model training. In addition, the data had to be transformed such that each data point had the same input size. Each raw data point represents a time series of different length, due to the different duration of each measurement, and of dimensionality 9 (i.e., 9 input features characterizing each movement, as explained in the previous section). Therefore, the data had to be reshaped. Two approaches were tested. First, the data was padded, i.e., zero values were attached to each of the 9 input vectors such that they had the same length. In a second approach, we tested linear interpolation as implemented in numpy [26] to achieve a fixed length for each vector. Each data point then consisted of 9 vectors, each of length 898, summing up to 8,082 values per data point, representing one tumbling element performed by one athlete. This data was then normalized in each direction, concatenated and reduced in its dimensionality for visualization to get a first insight into how the data is clustered and what could be more suitable for ML algorithms. This data is referred to as "raw data".

For dimensionality reduction, we performed principal component analysis (PCA) [27] in combination with *k*-means clustering [28] using scikit-learn [29]. The principal components plotted against each other for each data series, as well as their explained variance ratio, can be found in Figure A1 in the Appendix. For comparison, the data was further transformed into a power spectrum via an autocorrelation function using fast Fourier transform (FFT) [30,31] and the Wiener-Khinchin theorem [32]. The data was further interpolated to 1,000 data points per direction. We refer to these data as *spectra data*. The dataset was partitioned into a holdout test set containing 254 data points and a training set comprising 848 data points. To ensure proper learning and generalization to athletes besides those in the training set, the holdout test set contained 4 athletes that were not included in the training set.

2.4.2. Classification and model training

To learn the relationship between the time-series data and specific tumbling elements, supervised ML classification algorithms were employed [33]. These algorithms were trained to accurately map the input features to the corresponding tumbling elements. Gaussian Process Classification (GPC) [34,35], as implemented in the scikit-learn library [29], was selected for its robustness within small to medium-sized data and allowing for the prediction of an uncertainty measure next to each predicted class [36]. Training a GPC model involves learning the hyperparameters of the kernel function by maximizing the marginal likelihood of the training data [34]. This process adjusts the kernel parameters to fit the data while balancing model complexity and overfitting [36]. Once trained, the model computes predictions for unseen data by conditioning the Gaussian process on the observed training data and applying Bayes' rule [37]. The result is a predictive distribution for the latent function at each test point, which is then mapped to class probabilities [35]. Various GPC models were evaluated using different kernel combinations to determine the most effective configuration for classification tasks (see Table A1 in the Appendix). Among the tested kernels, the product of a Constant Kernel and a Rational Quadratic Kernel [38] demonstrated superior performance, achieving the highest accuracy in identifying tumbling elements. The final method including data preprocessing and ML prediction can be seen in Figure 4.

To ensure the reliability and generalizability of the models, a stratified [39] group *K*-fold cross-validation approach [40,41] with 5 folds was utilized next to a standard *K*-fold cross-validation approach. The *K*-fold cross-validation approach can be seen in Figure 5 with the stratified group *K*-fold approach visualized in Figure A4 in the Appendix. This method maintained the distribution of tumbling elements across each fold, thereby preserving the integrity of the evaluation process. Therefore, the data set is split into training (blue) and validation (orange). In each round, the model is trained on the training set and evaluated against the validation set. The mean accuracy and standard deviation of the 5 models were calculated across the five folds to assess model performance comprehensively. Additionally, a randomized hyperparameter search [42] was conducted over 10 iterations to optimize the model parameters. The search aimed to identify the best-performing model based on the highest mean cross-validation accuracy. Once the optimal hyperparameters were determined, the best model was refitted using the entire training dataset and was tested against the holdout test set. The final model parameters are specified in Table 1.

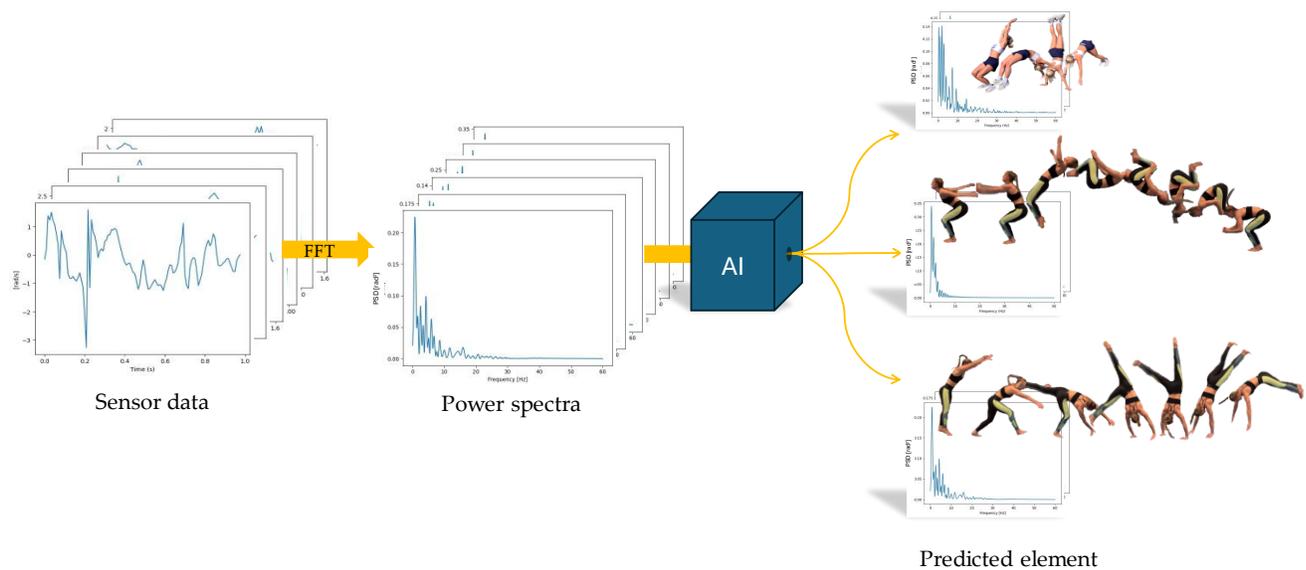

**Figure 4.** Data processing methodology showing raw data, spectra and machine learning classification of three different elements (from top: BHS, BT and RO).

**Table 1.** Summary of model parameters for different models found after randomized search.

| Data type | Split method | Constant value | α | Length scale |
|---|---|---|---|---|
| Raw Data | K-fold | 2205.374 | 0.129 | 2.662 |
| | Stratified Group K-fold | 9030.786 | 0.115 | 3.647 |
| Power Spectra | K-fold | 925.599 | 23618.319 | 22.788 |
| | Stratified Group K-fold | 116.713 | 0.752 | 0.291 |

To further understand the model's ability to learn from the data, learning curves were generated to evaluate how the performance changes as the size of the training dataset increases [43]. Learning curves provide insight into whether the model is underfitting or overfitting and how effectively it generalizes to unseen data. As the number of training samples increases, the model error on a holdout test set is expected to decrease and the accuracy is expected to increase. This behaviour can be examined by creating a learning-curve [43-45].

For this purpose, the training data was shuffled, and subsets of increasing size (100, 200, 400, 600 entries) were extracted. For each subset, the model was refitted using (stratified group) K-fold cross-validation (5 folds), ensuring that the distribution of tumbling elements was maintained across all folds. After training on each subset, the model's performance was evaluated by predicting the tumbling elements in the holdout test-set. The scoring metric used was the mean accuracy of the model on the holdout set, averaged across the five folds, along with the standard deviation between the folds.

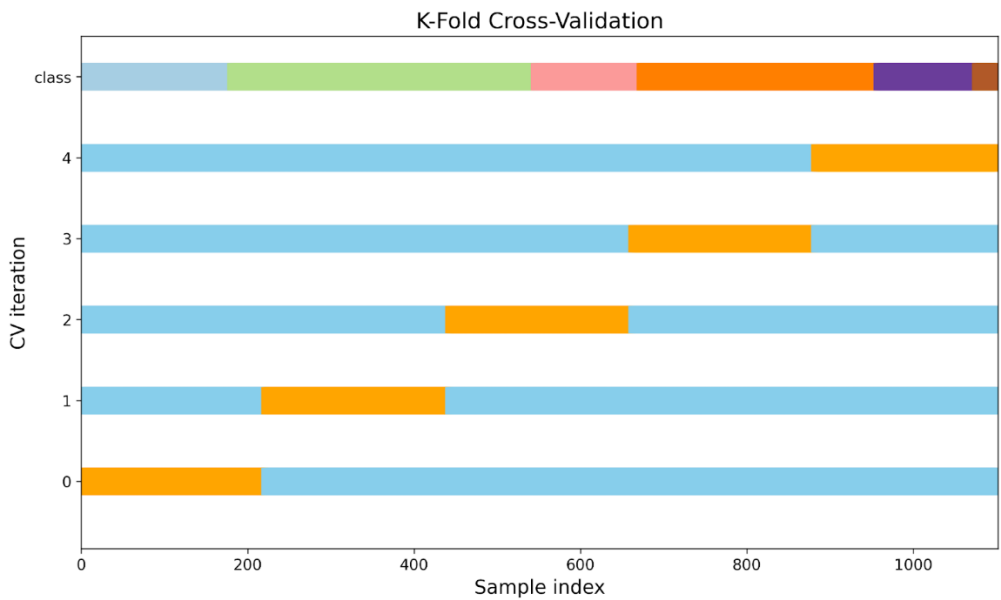

**Figure 5.** Schematic representation of K-fold cross-validation (CV) using five folds (*K*=5).

2.4.3. Model analysis

To better understand the contributions of individual features to the classification of tumbling elements, a feature importance analysis [46,47] was conducted. This analysis provides insight into which input features, such as triaxial accelerations or rotational speeds, are most influential in distinguishing between tumbling manoeuvres. Specifically, feature importance was evaluated by analysing the impact of perturbing or removing individual features on the model's performance. This was achieved through a permutation importance approach [47], where feature values within the test set were permuted while keeping other features intact, and the corresponding decrease in model accuracy was measured [46-48].

## 3. Results

First, the raw data is interpolated to the same size as each measurement was conducted for a different duration. PCA results (Figure A1), such as the adjusted rand index [49], justify the use of interpolated data in comparison to padded data, led to higher prediction accuracy when conducted employing initial ML tests. The duration of the different tumbling elements shows high variance between the elements but also between different athletes performing the same element (Figure A2). Even when looking at the same element this variance can be observed, making it necessary to extract important features to a time independent scale (Figure A3). Therefore, we transferred the acceleration data into the frequency domain using the FFT of the autocorrelation function [30-32], to ensure the model learns the features themselves instead of recognizing the length difference between exercises and increasing comparability between repetitions. This step was essential to extract meaningful frequency components from the data, such as dominant frequencies and spectral energy distributions, which are highly relevant for distinguishing between different tumbling elements.

The frequency-domain representations of the data were then used as inputs for the ML model, i.e., GPC models. By leveraging the frequency-domain features, the model could focus on periodic patterns and spectral characteristics that are not immediately apparent in the raw time-domain signals. The GPC model, trained on these frequency-domain features, was optimized to classify tumbling elements with high accuracy. Using this approach, we achieved an accuracy of 87.8% and 88.6% when using spectra data as input to GPC models with *K*-fold cross validation and stratified group K-fold cross validation, respectively. In comparison, using raw data without FFT led to an accuracy of ML models of 77.6% when using *K*-fold cross validation and stratified group K-fold cross validation. The resulting confusion matrix illustrates the amount of correctly predicted tumbling elements (diagonal) and wrongly predicted tumbling elements (off-diagonal).

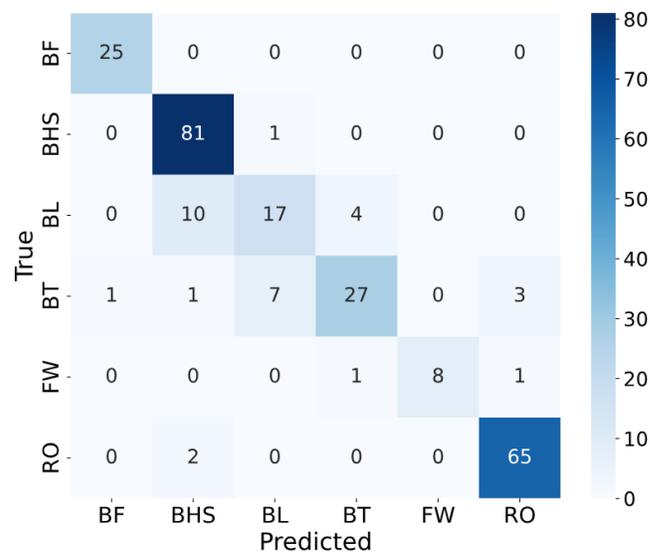

**Figure 6.** Confusion matrix showing correctly and wrongly predicted tumbling elements by the Gaussian Process Classification (GPC) model. Abbreviations of tumbling elements are the same as in Figure 2.

To ensure that models are learning properly, we further conducted learning curves (Figure 7a) and analysed the importance of the different features used as ML inputs (Figure 7b). The learning curves show how the model increases in accuracy with increasing data set size, while feature importance analysis is crucial for interpreting model behaviour and identifying the most relevant biomechanical signals that drive the GPC model. Comparison of a learning curve using Stratified group K-fold Cross validation (KF CV) is illustrated in Figure 7 a) for comparison.

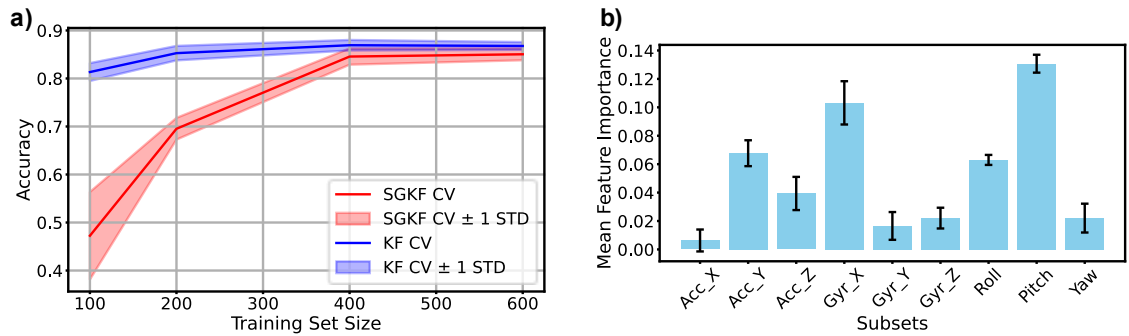

**Figure 7. a)** Learning curves of the final Gaussian Process Classification (GPC) model using K-fold (KF) cross validation (CV) Stratified Group (SG) KF CV with 5 folds. The testing is conducted on a hold-out test set containing athletes not included in the training and validation set. **b)** Feature importance of the input features used for training the GPC model using 5-fold CV.

## 4. Discussion

*4.1. Learning of the classification model*

As illustrated in the results section, the GPC model can predict tumbling elements with high accuracy. We performed analysis using classic K-fold cross validation (KF CV in Figure 7a) and compared it to stratified group KF CV, which allowed a clear distribution of the different athletes and tumbling elements, ensuring that no data leakage is present. In KF CV, the dataset is randomly split into K folds, with each fold used as a validation set once while the remaining K-1 folds are used for training (Figure A2a). However, in datasets where multiple samples originate from the same athlete (e.g., different tumbling elements performed by the same individual), this approach risks data leakage [50]. Specifically, data from the same athlete could appear in both the training and validation sets, allowing the model to learn individual-specific patterns rather than generalizable features, thereby overestimating performance. To address this, we employed SGKF CV (Fig. S4b and c), which groups samples based on athletes, ensuring that no athlete's data appears in both the training and validation sets. This approach prevents data leakage and ensures that the model is evaluated on entirely unseen athletes, providing a more realistic estimate of its generalization capability. Additionally, SGKF CV incorporates stratification to maintain the proportional distribution of tumbling elements across all folds, ensuring balanced representation of each class in both training and validation sets. Importantly, the CV is conducted to find ideal model hyperparameters. For testing, i.e., accuracy values shown in the plot, we used a hold-out test set that was not included in the training and validation set used for the learning curve. As can be seen, both methods show similar results, leading to the conclusion that data leakage is not an issue here, further showing that the models do not show a "Clever Hans" effect [20] and can generalize well to unseen data. For larger training set sizes (400 to 600 entries), both curves converge to similar accuracy values of approximately 0.9, with a reduced standard deviation, indicating improved model stability and generalization as more training data is included. However, at smaller training set sizes, the SGKF CV curve (red) exhibits a much higher error and variance compared to the KF CV curve (blue).

Specifically, the accuracy for SGKF CV drops significantly for a training size of 100, and the variance (±1 standard deviation) is also notably larger. This discrepancy is likely due to the stricter constraints imposed by SGKF CV, which ensures that entire groups (e.g., data from specific athletes) are kept separate between training and validation folds. In small datasets, this grouping constraint reduces the diversity of the training data, as fewer groups are available in each fold. Consequently, the model struggles to generalize effectively, leading to poor performance on the holdout set. As both methods led to similar results, further analysis is carried out using the 5-fold CV trained model.

As already mentioned, feature importance analysis was carried out to identify the key features contributing to the classification of tumbling elements. However, it is important to note that the features used in this study are not entirely independent of one another. For example, rotational speed about the *x* axis (Gyr_X) is correlated with rolling movements, i.e. rotations about the transverse body axis. Similarly, Gyr_Y is associated with rotations about the pitch axis, reflecting rotations about the longitudinal axis. Gyr_Z corresponds to angular velocity around the yaw axis, capturing rotations about the frontal body axis. While these features are not linearly dependent, they are biomechanically related as they collectively describe the complex rotational dynamics of tumbling elements. Despite this interdependence, feature importance analysis remains a valuable tool. Permutation-based approaches account for the combined effects of features in the model by measuring the incremental impact of perturbing individual features while retaining their correlations with others [46]. In this context, feature importance reflects the relative contribution of each feature to the model's decision-making process within the existing feature space. While Gyr_X, Gyr_Y, and Gyr_Z are related to pitch, roll, and yaw movements, respectively, their individual importance scores highlight the specific rotational axes that play the most significant roles in distinguishing tumbling elements in cheer sports.

The results of the analysis, shown in Figure 7b, revealed that certain features consistently contributed more to the model's performance. Specifically, rotational speed along the pitch and roll axes (Gyr_Y and Gyr_X) demonstrated the highest importance, underscoring their relevance in capturing the dynamic angular motions characteristic of tumbling manoeuvres. These rotational features reflect key biomechanical elements, such as body orientation and twisting motions, which are critical for differentiating between tumbling elements. Vertical acceleration (Acc_Z) also emerged as highly influential, highlighting its role in capturing ground reaction forces and vertical displacements during tumbling. Conversely, yaw rotations (Gyr_Z) and horizontal accelerations exhibited lower importance, suggesting they were less critical for classification in this dataset. These results align with the biomechanical understanding that the tumbling elements studied are predominantly characterized by characteristic rotations about the transverse and frontal axes, while rotations about the longitudinal body axis are usually part of a combined set of rotations with the transverse and frontal axes being more decisive for characterization.

*4.2. Model analysis with respect to cheerleading elements and their correlations*

Finally, the confusion matrix shows the different elements that were predicted correctly (diagonal elements) within the test set and that were confused with each other (off-diagonal elements).

The model demonstrated strong classification performance for BHS and RO, with 81 and 65 true positives, respectively, and minimal misclassifications. This indicates that these elements have distinct biomechanical characteristics captured effectively by the model's feature space. Similarly, the model showed reasonable performance for BT with 28 true positives, though there were some confusions with BL and RO, which share overlapping features in their execution.

For the BL class, the model achieved moderate accuracy with 18 true positives but exhibited significant misclassifications. Specifically, BL was misclassified as BHS in 9 instances and as BT in 7 instances. This confusion is likely due to similarities in the rotational dynamics and most probably further exacerbated by inaccurate body position execution of the BL and BT elements by several individual athletes: Both elements involve an aerial full 360° backward rotation about the athlete's lateral axis without any ground contact in the flight phase [3]. The BHS, in contrast, is distinct by its short handstand support phase at a backward rotation angle around the lateral axis of approximately 180°. Although this unique acceleration impact peak of a BHS should be, in principle, easy to detect from IMU data, some practical limitations might come into play. We used a single IMU attached to the athlete's lower back, which might not always be completely excluded from relative motion during vigorous movements. A certain degree of accelerometer measurement inaccuracy may thus be a factor. The confusion between BL and BT is most likely due to the similarity of the movement sequences, which differ only in the position of the legs in terms of the hip flexion angle [3], while the backflip rotations are similar (despite moderately differing moments of inertia about the lateral axis). The differentiation between BL and BT is further complicated when the execution of the BT is imprecise, as it was observed in several athletes. A significant share of athletes tended to adopt a hip-flexed pike-like position rather than the correct, fully extended layout position. Similarly, distinguishing between BT and a Back Pike, even though not required in this study, may be even more challenging.

The FW class, with 8 true positives, showed higher variability in predictions, likely due to its limited representation in the dataset. In the FW, there is a characteristic 360° forward rotation about the lateral axis of the athlete [3], which makes a confusion with a backflip element (i.e., BT and BL) unlikely. A FW has some characteristic kinematic similarities with an RO, but only up to about the middle of the first handstand phase. In contrast to the FW, the RO is then followed by an additional rotation around the body's longitudinal axis [3], which is absent in the FW. In general, smaller class sizes, such as those for BF and FW, contributed to reduced performance for these elements, likely due to insufficient training samples. This imbalance can limit the model's ability to generalize for underrepresented classes.

The analysis suggests several opportunities for improving classification performance. First, feature refinement could play a crucial role in enhancing separability between similar classes, such as BL and BT. Introducing advanced features, such as energy profiles, timing-based parameters, or transformations like wavelet analysis, could help better differentiate overlapping movements. Second, data augmentation and balancing techniques could address the underrepresentation of certain classes, such as BF and FW, to improve the robustness of predictions.

Nevertheless, the model shows that similar elements that were not conducted cleanly or executed correctly are more likely to be misclassified, leaving room for subjective interpretation and judging of elements. This observation highlights the potential of integrating judging criteria into the training process to better align the model's classifications with human judgments of quality and execution. By incorporating subjective scoring metrics, such as deductions for improper form, synchronization, or landing stability, future models could assist judges by providing objective and consistent metrics while leaving room for human interpretation in cases of ambiguity.

Such an integration would allow the model to move beyond mere classification of tumbling elements and assist in performance evaluation, potentially offering real-time feedback during competitions. For example, features related to execution quality could be trained alongside element classification, enabling the model to flag incomplete or incorrectly executed elements for further evaluation by judges. This could enhance the fairness and accuracy of scoring while reducing the cognitive load on human judges.

## 5. Conclusions

This study shows that machine learning can objectively classify tumbling elements in cheerleading using data from a single IMU. In total, 1102 recorded elements from 16 athletes, triaxial acceleration and rotational speed data were transformed into the frequency domain for improved classification. Gaussian Process Classification achieved high accuracy, generalizing well to unseen athletes. While challenges remain in distinguishing similar manoeuvres, these results highlight the potential of artificial intelligence to complement traditional judging. Future work should focus on refining feature extraction, expanding datasets, and integrating models that assess not only element classification but also execution quality and technical precision, ultimately supporting more objective and consistent scoring in competition and training.

**Author Contributions:** Conceptualization, OUe and JW.; methodology, OUe, JW; software, machine learning and data analysis, SW, RC, OUe, JW; validation, OUe, JW; formal analysis, EH, OUe; investigation, EH, OUe; resources, OUe, JW; data curation, EH; writing—original draft preparation, JW, OUe.; writing—review and editing, all authors; visualization, SW, EH, RC, OUe, JW; supervision, OUe, JW; project administration, OUe, JW, SP; funding acquisition, OUe, JW, SP. All authors have read and agreed to the published version of the manuscript.

**Funding:** This research was funded by the German Science Foundation (Deutsche Forschungsgemeinschaft, DFG), grant number 545264300. The APC was funded by the Magdeburg-Stendal University of Applied Sciences.

**Institutional Review Board Statement:** The study was conducted in accordance with the Declaration of Helsinki, and approved by both the Ethics Committee of the Department of Engineering and Industrial Design of the Magdeburg-Stendal University of Applied Sciences (protocol code EKIWID-2025-02-001EH, date 2025-02-07) and the Institutional Review Board of the University of North Florida (protocol code IRB#1970666-2, date 2023-01-18).

**Informed Consent Statement:** Informed consent was obtained from all subjects involved in the study.
**Data Availability Statement:** Data will be available upon request and an example of a few data points (pseudonymized) is uploaded with the code.

**Code Availability Statement:** The code used to produce the results and train the ML models can be found on figshare. https://doi.org/10.6084/m9.figshare.28303298.v2

**Acknowledgments:** The authors wish to thank all participants of this study for their commitment. We further thank Birte Scholz and Niklas Meier for proofreading our manuscript and providing feedback.

**Conflicts of Interest:** The authors declare no conflicts of interest.

## Appendix A: Machine learning

*A1. Data pre-processing*

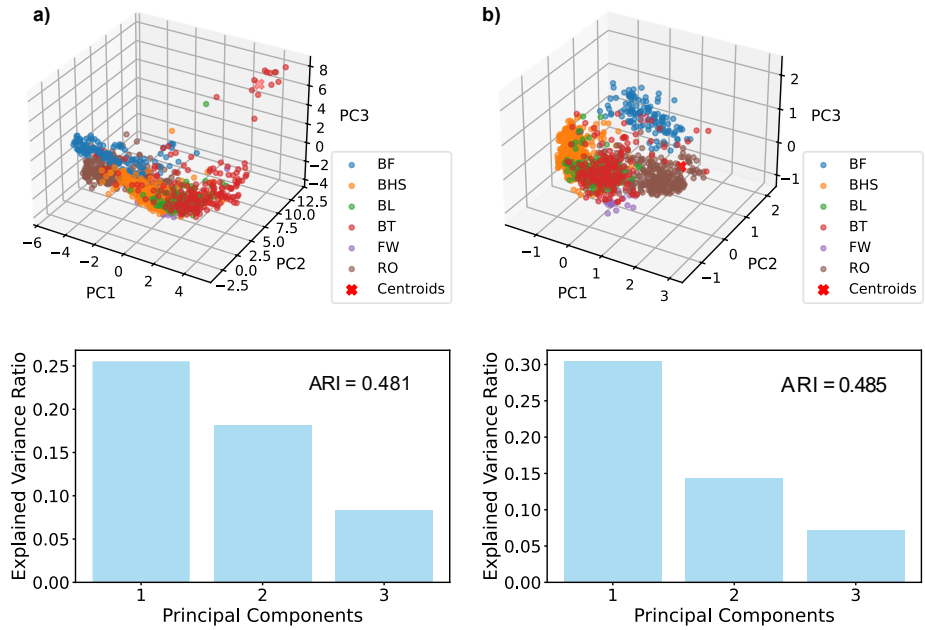

**Figure A1.** PCA plots with k-means clustering (top) and explained variance ratio of these PCA plots as well as the adjusted Rand Index (ARI) for the clustering (bottom). **a)** Results for padded spectra data, and **b)** interpolated spectra data.

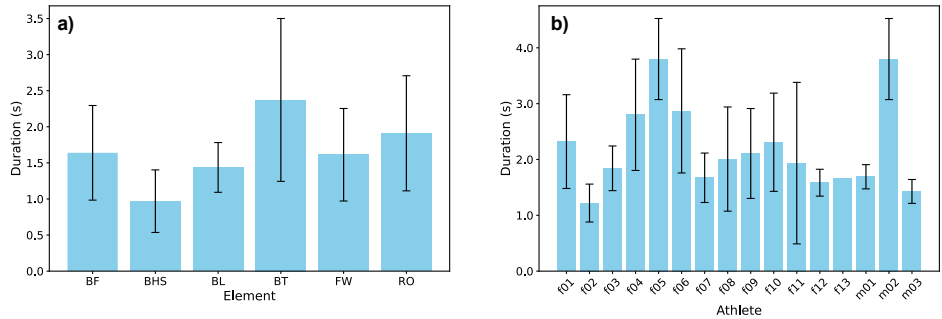

**Figure A2.** Comparison of mean durations of tumbling elements (error bars: SD). **a)** Mean durations of the different elements over all repetitions. **b)** Mean durations of the BT element performed by different athletes.

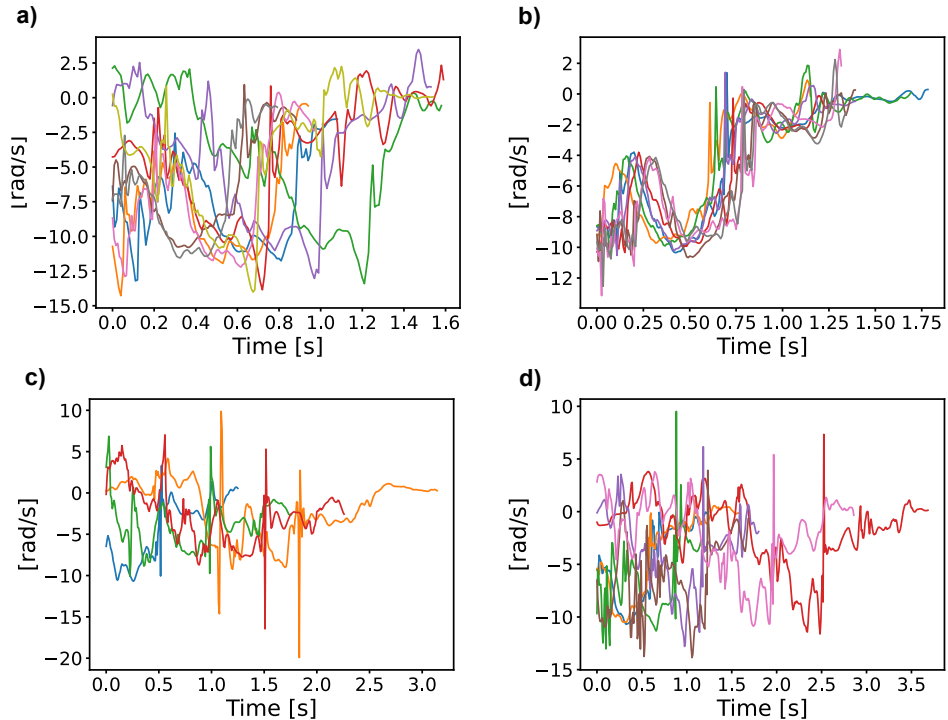

**Figure A3:** Raw Gyr_X data of four different athletes (depicted in (a) through d)) performing BT. The duration of this element varies between repetitions and athletes with an increasing maximal duration from a) to d).

*A2. Cross-validation of classification models*

Figure A4a provides a schematic representation of the different cross-validation (CV) techniques used in this study to evaluate the performance of the Gaussian Process Classification (GPC) models. Cross-validation is a critical step in ensuring the reliability and generalizability of a model by partitioning the dataset into distinct training and validation sets. The methods illustrated here include standard K-fold cross-validation (a), stratified K-fold cross-validation (b), and grouped *K*-fold cross-validation (c), each tailored to address specific characteristics of the dataset and classification task. In this study, we used (a) and a combination of (b)+(c).

In standard K-fold cross-validation (Figure A4a), the dataset is split into K equally sized folds. Each fold is used once as the validation set (orange segments), while the remaining K-1 folds serve as the training set (blue segments). This method does not account for class distribution or group dependencies, making it suitable for datasets without structural or hierarchical groupings.

Stratified *K*-fold cross-validation (Figure A4a) ensures that each fold maintains the same proportion of class labels as the original dataset. This approach is especially useful for imbalanced datasets, where preserving class distributions during training and validation is crucial to avoid biased performance evaluations. The stratified sampling ensures that each fold is representative of the overall data distribution.

Grouped *K*-fold cross-validation (Figure A4c) addresses scenarios where individual data points are not independent but instead belong to identifiable groups (e.g., individual athletes or sequences in cheerleading routines). In this method, groups are kept intact, and no group appears in both training and validation sets simultaneously. This prevents data leakage and ensures that the model is evaluated on entirely unseen groups, reflecting its true generalization ability.

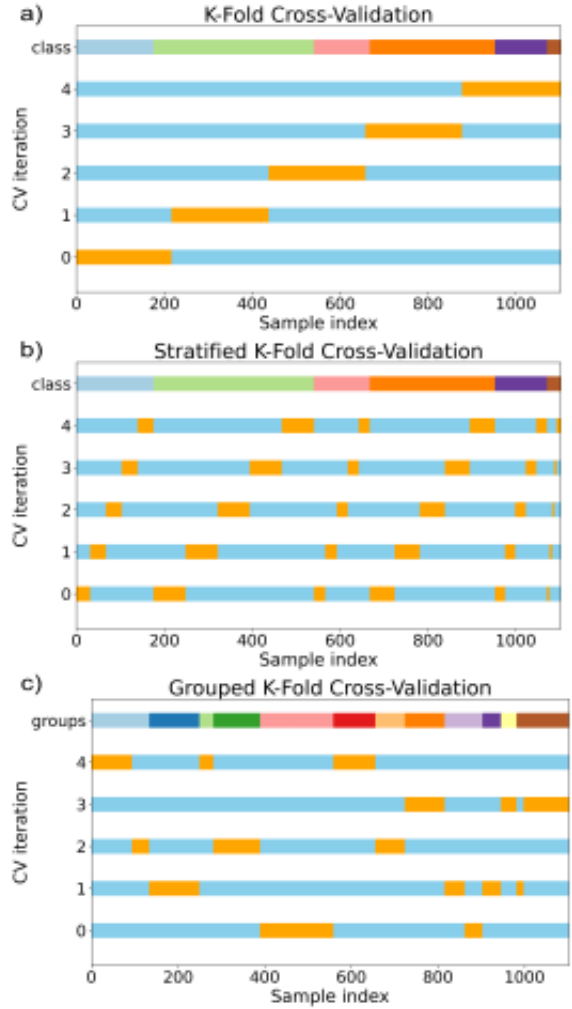

**Figure A4:** Schematic representation of cross-validation techniques used in this study. **(a)** Standard K-fold cross-validation, where the dataset is randomly split into *K* folds, with one-fold serving as the validation set (orange) and the remaining folds as the training set (blue). **(b)** Stratified K-fold cross-validation, which ensures the class distribution in each fold reflects the overall dataset. **(c)** Grouped *K*-fold cross-validation, where data points belonging to the same group (e.g., individual athletes) are kept together, ensuring no group appears in both the training and validation sets

*A3. Different kernel performances*

**Table S1.** Comparison of 6 different kernel Combinations always consisting of a KOnstant Kernel C combined with a second kernel. Models were trained on spectral data with stratified group k-fold cross validation during Hyperparameter optimization and evaluated on the holdout test set.

| Kernel Combination | C*Matern | C*RBF | C*Rational quadratic | C+Matern | C+RBF | C+Rational Quadratic |
|---|---|---|---|---|---|---|
| Test accuracy | 88.50% | 87.80% | 88.60% | 83.10% | 84.30% | 84.30% |